\documentclass[12pt]{iopart}

\usepackage{xcolor, graphicx}
\usepackage[normalem]{ulem}
\usepackage{braket}

\newcommand{\void}[1]{}

\newcommand{\be}{\begin{equation}}
\newcommand{\ee}{\end{equation}}
\newcommand{\ba}{\begin{align}}
\newcommand{\ea}{\end{align}}

\begin{document}
\title{Coupled coherent states method for
tunneling dynamics: an interpretative study}

\author{F. Grossmann}
\address{Institut f\"ur Theoretische Physik, Technische Universit\"at Dresden, D-01062 Dresden, Germany}

\ead{frank@physik.tu-dresden.de}

\date{\today}

\begin{abstract}
Numerical solutions of the time-dependent Schr\"odinger
equation based on the variational principle may offer
physical insight that cannot be gained by a solution
using fixed grids in position and momentum space.
Here we focus on the tunneling dynamics in a quartic
double-well and the use of classical,
trajectory-guided coherent states
to gain insight into the workings of the coupled coherent states method developed by Shalashilin and Child
[J. Chem. Phys. {\bf 113}, 10028 (2000)].
It is shown that over-the-barrier classical trajectories,
alone, can accurately describe the tunneling effect.
\end{abstract}

\maketitle

\section{Introduction}

Since several decades, variational methods to solve the
time-dependent Schr\"odinger equation (TDSE) have been
used in the chemical physics community
\cite{BJWM00,Lubi,Richings2015,irpc21}, where they are employed to
study the quantum dynamics of molecular
systems with a considerable number of degrees of freedom. Although
seminal work on the variational principle is also known in the theoretical physics community \cite{KS81}, related numerical approaches
in the time-domain have seen widespread use only more recently, with the advent of matrix product state techniques \cite{Hetal11,Scholl2011} for the study of entanglement in finite lattice dynamics.

In molecular physics, the use of Glauber coherent states
\cite{Glauber} with time-dependent parameters is especially appealing, because (i) often the initial states
are ground vibrational states of the ground electronic potential energy surface and thus of Gaussian nature and (ii) methods based on Gaussian basis functions allow to gain an
understanding of the quantum dynamics by, e.g., restricting the
center parameters to follow classical trajectories in phase space. An
intriguing example is given by a study on the one-dimensional Morse oscillator model that has been done
by Wang and Heller. By using the semiclassical Herman-Kluk (HK)
propagator \cite{HK84}, these authors have shown that the quantal
revival dynamics present in this system can be
understood as a subtle interference phenomenon
of classical trajectories that are spread ``all over phase space''
\cite{WaHe09}. Whereas the HK propagator has also proven useful in the
study of Bose-Hubbard dynamics for small well numbers
\cite{jpa16,SiSt14,Letal19},
its use for the tunneling dynamics in a double well is
not appropriate, if a single wavepacket calculation is desired \cite{SC01}.
We will therefore base our numerical investigations on the so-called
coupled coherent states (CCS) method by Shalashilin and Child \cite{SC00}. In this method, the coherent state parameters are still classical, like in the HK case, but the coefficients in the
expansion of the wavefunction are determined variationally,
in contrast to the HK case, where they again follow
from the (linearized) classical dynamics, see \cite{irpc21}
for a recent review.

The presentation is structured as follows: First in Section \ref{sec:dwell},
we briefly review the Hamiltonian of the quartic double well.
In Section \ref{sec:CCS} we then recapitulate the coupled coherent states method \cite{SC00}
and the modification of the classical Hamiltonian
due to normal ordering.
In Section \ref{sec:num} we present numerical results
for the tunneling dynamics, using suitably tailored rectangular
grids of initial conditions in phase space, that allow us
to discriminate the power of low as well as of
high energy classical trajectories to describe tunneling.
A brief summary and outlook are given in the last section.

\section{The quartic double-well model}\label{sec:dwell}

We start the main body of the presentation with a brief review of the 
double-well potential in one dimension, which is a model well suited
to study coherent tunneling dynamics. It has many applications, one of
the most important ones being the ammonia molecule, first
discussed with respect to the tunneling effect by Hund as early
as 1927 \cite{Hu27}. A solid state realization is given by a suitably parametrized rf-SQUID, where the role of the coordinate is played by the flux through the ring \cite{Ku72}. More recently bistable potentials have been discussed in cold-atom physics in connection with Bose-Einstein condensation \cite{Ketal08}. The effect of an external
sinusoidal field on the tunneling is quite counterintuitive,
as an appropriately chosen field can lead to a complete stand-still of the tunneling dynamics \cite{zpb91}.

For a particle of unit mass, the quartic double-well oscillator
is governed by a Hamilton operator of the form
\be
H(\hat p,\hat q)=\frac{\hat p^2}{2}+V(\hat q),
\label{eq:dw}
\ee
with the potential function
\be
V(q)=-\frac{a}{2}q^{2}+\frac{b}{4}q^{4}+E_B,\qquad
a, b\in \Re^{+} .
\ee
It has quadratic minima at $q_{\pm}=\pm\sqrt{\frac{a}{b}}$ and a quadratic maximum at $q_{0}=0$ of height $E_{B}=\frac{a^{2}}{4b}$.
By demanding that the second derivative of the potential
at the symmetric minima be unity and the barrier height be $D$,
we fix the parameters to be $a=1/2$ and $b=1/(16D)$.

\begin{figure}
\includegraphics[width=0.99\columnwidth,trim = 0cm 5cm 0cm 5cm, clip]{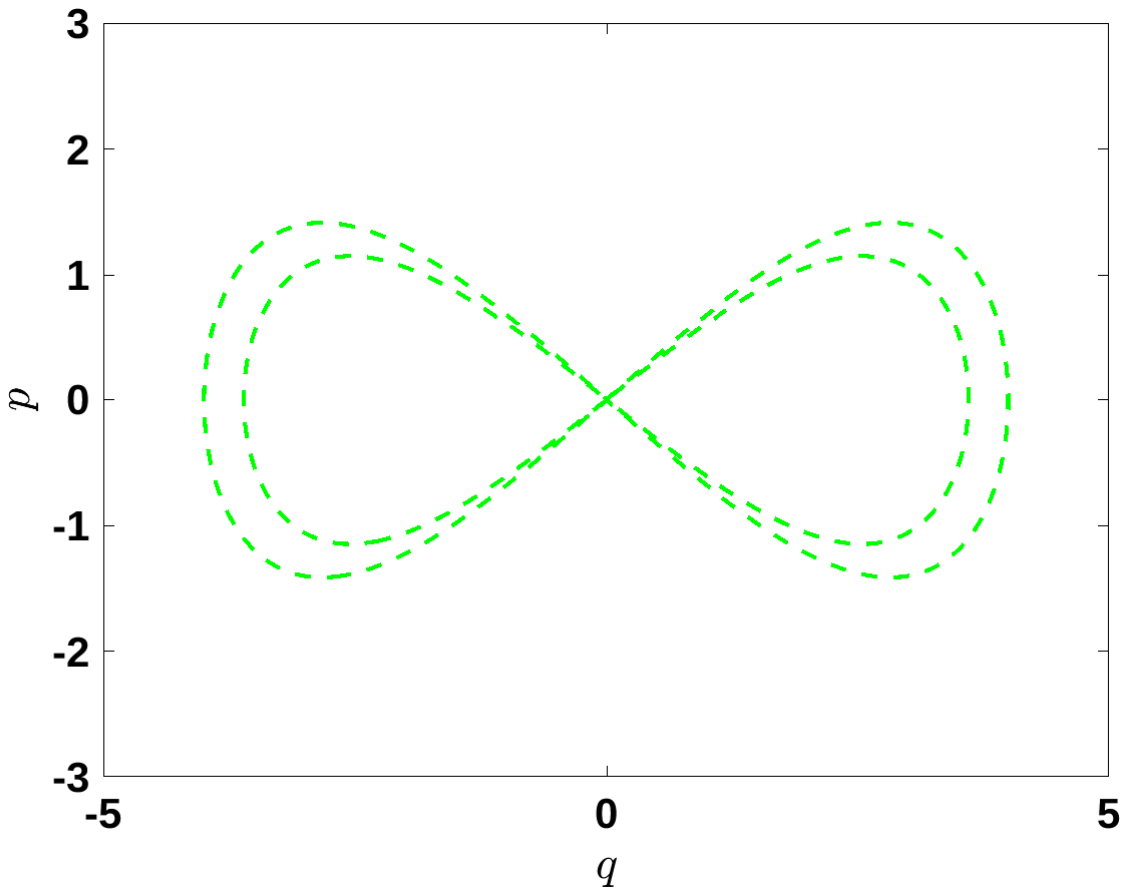}
\caption{Separatrix in the phase space of the quartic double well with $D=1$ for the standard classical Hamiltonian (the larger ``8'') and for its normal ordered version (the smaller ``8'').}
\label{fig:eight_1}
\end{figure}

The phase space portrait of the dynamics contains the prototypical separatrix, defined by $H(p,q)=D$ and shaped like the number eight rotated by 90 degrees, as displayed in Fig.\ \ref{fig:eight_1}, as well as a hyperbolic fixed point at $p=0,q=0$ on that separatrix and two elliptic fixed points at $p=0,q=q_{\pm}$ \cite{Reichl}. In the figure, we
also display the modification of the separatrix due to
normal ordering, as explained in Sect.\ \ref{sec:CCS}.
If the available energy is higher than the barrier, a classical
particle  starting on the right side of the barrier can reach the
left one and get back in a periodic fashion.

Quantum mechanically, in the usual tunneling scenario, where a Gaussian
is sitting at the minimum of one of the two wells initially, with
average energy below the barrier, the particle is still moving to the
other well and back with a (usually very small) frequency
given by the difference $E_2-E_1$ of the two lowest eigenvalues
of the Hamilton operator.
This motion is also referred to as umbrella motion. The case of
an initial state starting from the top of the barrier is referred
to as toppling pencil motion and has been studied classically in detail
by Dittrich and Pena Mart\'inez \cite{DPM20} as well as quantum
mechanically in \cite{pra22_1}, where the influence of a finite number
of bath degrees of freedom on the central degree of freedom (the double well) has been investigated.

\section{The method of coupled coherent states}
\label{sec:CCS}

The numerical method for the solution of the TDSE,
to be used in the following, is the
coupled coherent states method developed by Shalashilin and Child
\cite{SC00,ShBu08}. It is most clearly formulated
by using a complexified phase space variable, defined by
\be
z(t)=\frac{\gamma^{1/2}q(t)
+{\rm i}\hbar^{-1}\gamma^{-1/2}p(t)}{\sqrt 2},
\ee
where $\gamma=m\omega/\hbar$.
In the remainder of the presentation, we set
$\hbar$ as well as $m$ equal to unity and use Gaussian basis functions,
which correspond to the coherent states of a harmonic oscillator
with oscillation frequency $\omega=1$ (in dimensionless units) and
the definition above simplifies considerably.

In the coupled coherent states (CCS) method of Shalashilin and
Child, an ansatz is made for the time-evolved wavefunction as a
finite esum in the form
\begin{equation}
\label{eq:ansa}
 \Ket{\Psi(t)}=\sum_{l=1}^Ma_l(t)\Ket{z_l(t)},
\end{equation}
where each ket $\Ket{z_l(t)}$ is a coherent state (an eigenstate
of the annihilation operator) with a time-dependent, complex
center parameter $z_l(t)$ and the ``multiplicity'' of the ansatz
is given by $M$.
The set of the $M$ center parameters is following
uncoupled classical trajectories, each of which is
fulfilling \cite{ShBu08}
\be
\label{eq:EOM}
 {\rm i}\partial_tz=\partial_{z^\ast}H_{\rm ord}(z^\ast,z),
\ee
which are Hamilton's equations in complex notation.
The normal ordered Hamiltonian is gained by using
\be
\hat{q}=\frac{1}{\sqrt{2}}(\hat{a}^\dagger+\hat{a}),
\qquad
\hat{p}={\rm i}{\frac{1}{\sqrt 2}}(\hat{a}^\dagger-\hat{a}),
\ee
in Eq.\ (\ref{eq:dw}), and moving all creation operators to the left,
by using the fundamental commutation relation
\be
[\hat a,\hat a^\dagger]=\hat 1.
\ee
Then we replace $\hat a^\dagger$ by $z^\ast$
and $\hat a$ by $z$ to arrive at the classical Hamiltonian \cite{pra22_1}
\begin{eqnarray}
H_{\rm ord}&=&-\frac{1}{4}\left[\left(z^\ast-z\right)^{2}-1\right]
-\frac{1}{8}\left[\left(z^\ast+z\right)^{2}+1\right]
\nonumber
\\
&&+\frac{1}{256 D}
\left[\left(z^\ast+z\right)^{4}+6\left(z^\ast+z\right)^{2}+3\right].
\end{eqnarray}
We stress that by expressing the $z$ variable again in
terms of position and momentum and by gathering all constant terms in the potential, a modified, normal
ordered version of the potential emerges (see also
\cite{SC01}), that is given
by
\be
V_{\rm ord}(q)=D-\frac{1}{4}\left(q^{2}-\frac{1}{2}\right)
+\frac{1}{64D}\left(q^{4}+3q^2+\frac{3}{4}\right).
\ee
Its minima are slightly shifted to
$q_{\pm}=\pm\sqrt{8D-3/2}$ and the relative maximum at
$q=0$ is increased to $V_{\rm ord}(0)=D+1/8+3/(256D)$.
Due to the fact that the value of the ordered potential at
the mimima is now not zero any more but given by
$V_{\rm ord}(q_\pm)=1/2-6/(256D)$, the barrier height
is changed (typically decreased)  to $D-3/8+9/(256D)$. These changes also lead to a change of the separatrix, which is highlighted in Fig.\ \ref{fig:eight_1}.
It is especially important to note that now already
smaller values of phase space variables are leading to
a crossing of the barrier.

The equations of motion for the $a$-coefficients are derived from the time-dependent variational principle \cite{KS81,ShBu08} and are linear, coupled differential equations of the form
\begin{equation}\label{eq:ccs}
    {\rm i}\sum_{l=1}^M\langle z_k(t)|z_l(t)\rangle\dot{a}_l(t)=\sum_{l=1}^M{\tilde H}_{kl}(t) a_l(t),
\end{equation}
with the time-dependent matrix elements
%
\begin{eqnarray}
\label{eq:htil}
    {\tilde H}_{kl}(t)=\langle z_k(t)|z_l(t)\rangle
    \bigg[ H_{\rm ord}(z_k^\ast,z_l) &-& \frac{1}{2}\left(z_l(t) \frac{\partial H_{\rm ord}}{\partial z_l}-\frac{\partial H_{\rm ord}}{\partial z_l^\ast}z_l^\ast(t)\right)
    \nonumber\\
    &-&\left. z_k^\ast(t)\frac{\partial H_{\rm ord}}{\partial z_l^\ast} \right],
\end{eqnarray}
and where an overlap matrix element, defined as
\begin{equation}
\label{eq:over}
\Braket{z_k|z_l}={\rm e}^{-\frac{1}{2}(|z_k|^2+|z_l|^2)+z_k^\ast z_l}
\end{equation}
appears on both sides of the equation. We stress that, obviously,
it may not be cancelled because the multiplication in
Eq.\ (\ref{eq:htil}) is an element wise multiplication. For
the inversion of the overlap matrix
we thus employ a regularization in the form of the addition of a unit
matrix multiplied by $10^{-8}$. Furthermore, the partial derivatives
of the ordered Hamiltonian may be replaced by the left hand side
of the classical equation of motion, Eq.\ (\ref{eq:EOM}). We
found that there is only a marginal difference in the  performance of
the code due to this trick, however.

Without loss of generality, we will start the quantum dynamics
with a single one of the $a_l$ being
non-zero initially, such that $\Ket{\Psi(0)}=\Ket\alpha$ is
given as the ground state of the above mentioned harmonic
oscillator with $\omega=1$. Due to the
coupling of the equations for the coefficients, generically,
the population
of the other basis functions will take on non-zero values during propagation.

\section{Numerical results}
\label{sec:num}

Before we show numerical results for the solution of the TDSE, we first focus on their central ingredients, i.e., the classical trajectories. If the energy of all the classical trajectories lies
below the barrier, i.e., if all the initial conditions are
inside the separatrix (see Fig.\ \ref{fig:eight_1}) and on
the right side only (all have positive values of $q$), then they will stay there and tunneling cannot be described
properly because the classical trajectories, by being uncoupled, cannot overcome the barrier and at no instance of time a Gaussian will ``appear'' on the other side.

Two alternative scenarios can be imagined to overcome this limitation:
\begin{itemize}
\item All trajectories start on one side only but
the energies of some of them are larger than the barrier
\item Some (initially unpopulated) Gaussians are
also started (with zero $a$-coefficient) on the other
side but all trajectories have energies smaller than the
barrier
\end{itemize}

\begin{figure}
\includegraphics[width=0.46\columnwidth,trim = 0cm 4.6cm 0cm 5cm, clip]{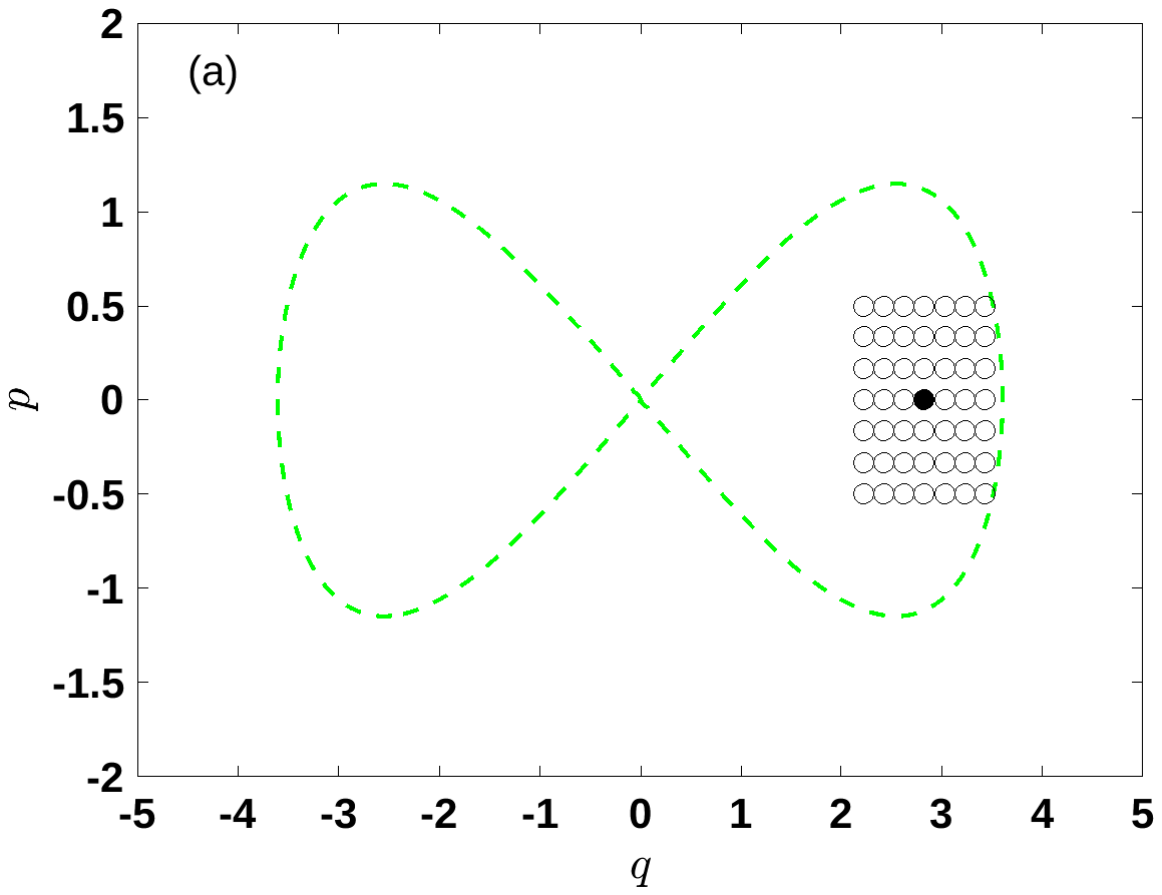}
\includegraphics[width=0.49\columnwidth,trim = 0cm 5cm 0cm 5cm, clip]{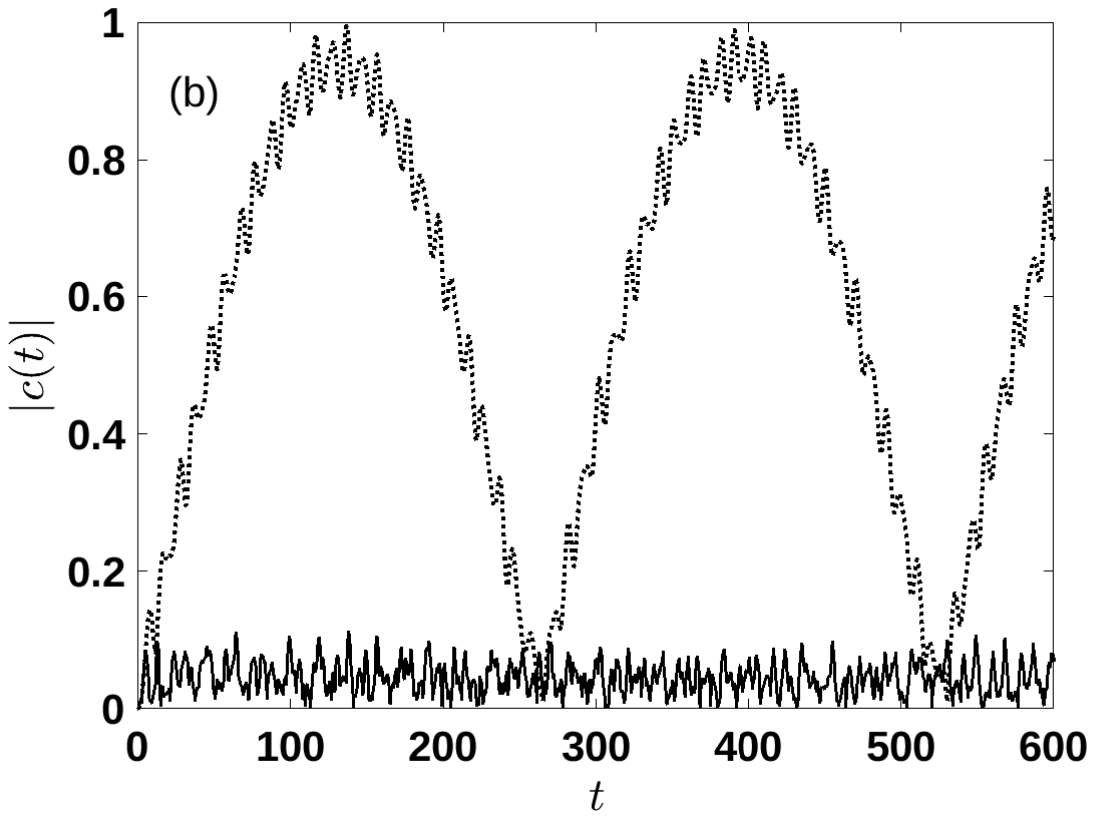}
\caption{(a) Separatrix of $H_{\rm ord}$ in the phase space
of the quartic double well with $D=1$ with unoccupied initial
conditions indicated by small open circles; the initially occupied
Gaussian is displayed by a filled circle.
(b) Comparison of numerical results for the absolute value of the
autocorrelation function, using a grid of $M=49$ initial conditions
shown in panel (a) in the CCS case (solid line) and converged split
operator FFT calculations (dotted line).}
\label{fig:eight_2}
\end{figure}

In order to discriminate the two cases in a clean
fashion, we refrain from a random sampling of initial conditions that has been used successfully (i.e., it was shown that CCS is capable of
correctly describing the tunneling effect) in \cite{SC01} and that
is especially favorable, if more than a single degree of freedom is to be investigated. With random samples
drawn, e.g., using an importance sampling procedure, the appearance of trajectories with larger energies (and on the other side of the barrier) cannot be avoided. We thus stick to rectangular grids, as displayed, e.g., in panel (a) of Fig.\ \ref{fig:eight_2}. The potential barrier's height, we study is given by $D=1$ and the initial wavefunction $\Ket \alpha$ is a Gaussian corresponding to the ground state of the harmonic approximation to the right well of the plain (not normal ordered) potential.

Firstly, we study the case of initial conditions displayed in
panel (a) of Fig.\ \ref{fig:eight_2}. We stress that only the
trajectory in the middle of the grid has a non-zero $a$ coefficient
at $t=0$. The time-dependent quantity, we monitor is the
cross-correlation function
\be
c(t)=\langle \beta |\Psi(t)\rangle=\sum_{l=1}^M a_l
\langle \beta|z_l(t)\rangle,
\ee
where $\beta=-\alpha=-\sqrt{8}$. From the tunneling splitting
\be
\Delta:=E_2-E_1\approx 2.392\times 10^{-2},
\ee
for $D=1$ given in \cite{zpb91}, we expect a tunneling period
of $T_t=2\pi/\Delta\approx 263$ (all in dimensionless units)
to show up in the time evolution of $c(t)$. In panel (b)
of Fig.\ \ref{fig:eight_2}, we see that the full quantum
solution, gained by a split-operator FFT implementation of the
TDSE \cite{Gross3} and displaying the predicted (long) period,
apart from the high-frequency, small-amplitude oscillations
(which are due to the presence of eigenstates, with energies
$E_n$ above the barrier ($n\geq 3$) in the initial state), is
not captured by our ansatz.

\begin{figure}
\includegraphics[width=0.46\columnwidth,trim = 0cm 4.6cm 0cm 5cm, clip]{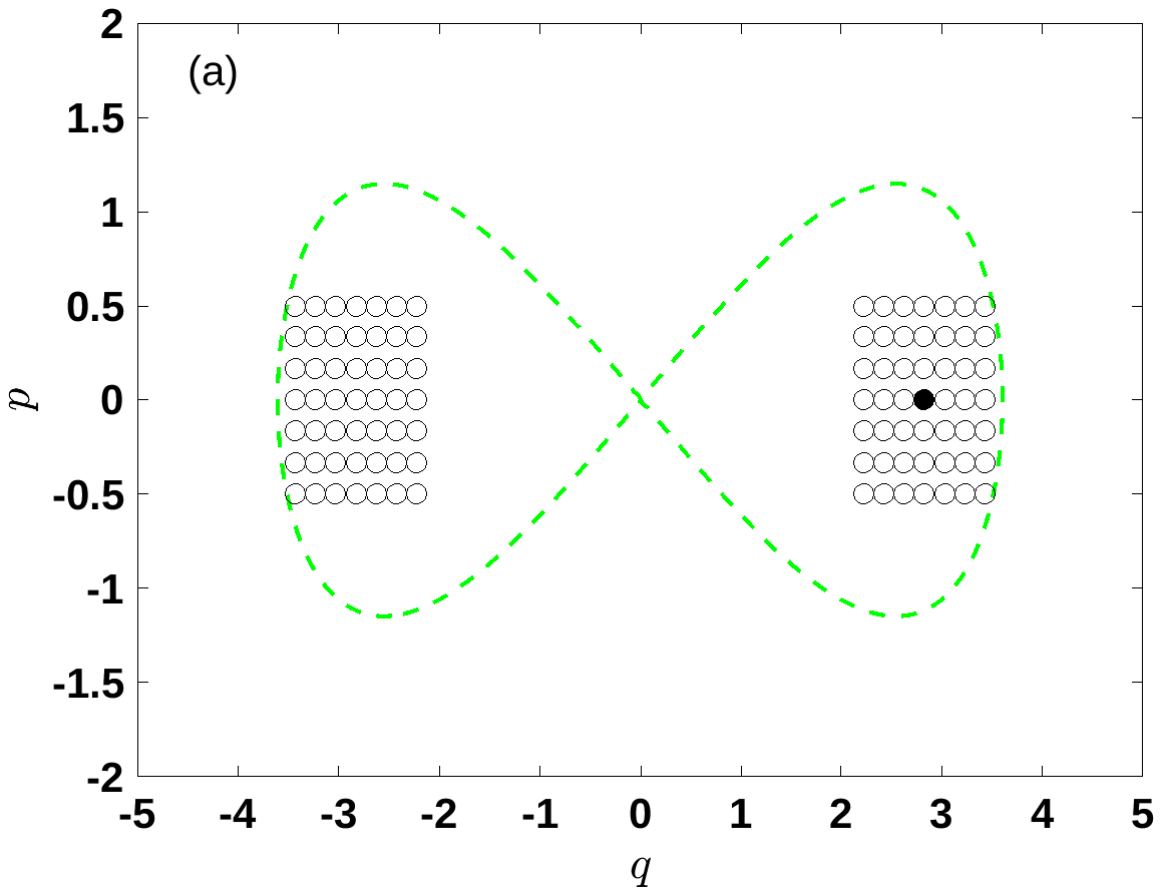}
\includegraphics[width=0.49\columnwidth,trim = 0cm 5cm 0cm 5cm, clip]{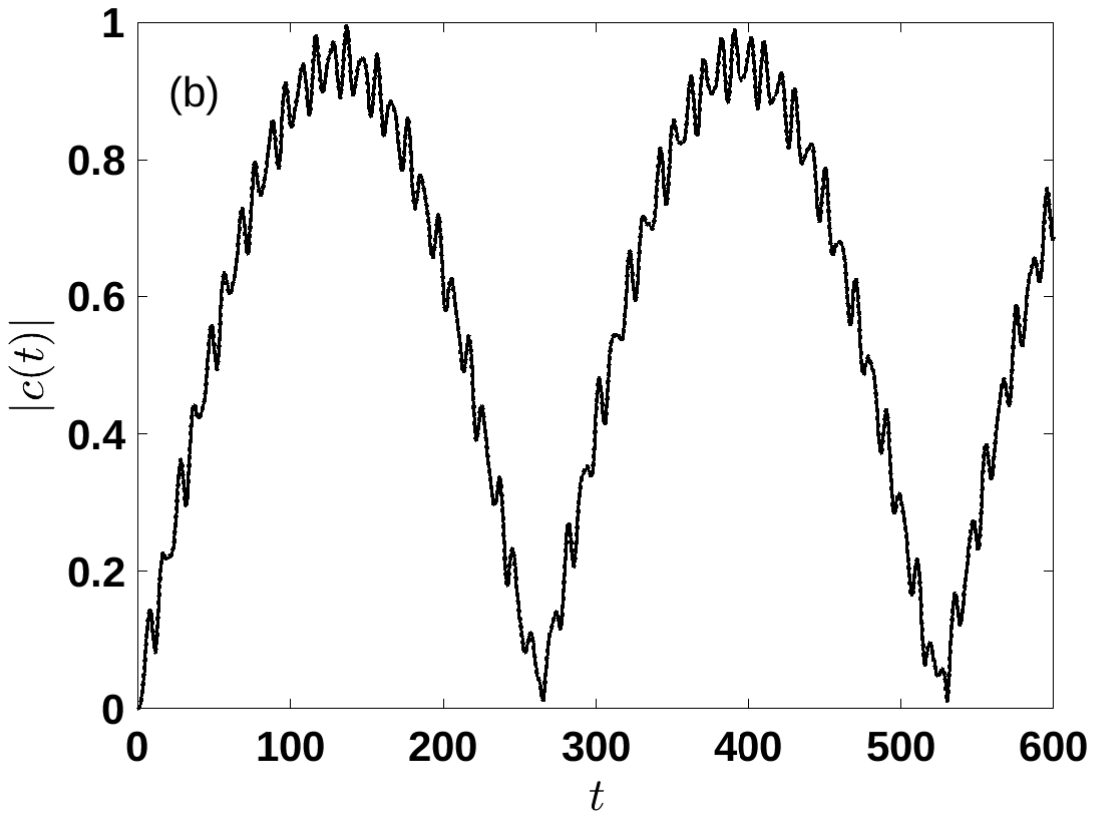}
\caption{(a) Separatrix of $H_{\rm ord}$ in the phase space
of the quartic double well with $D=1$. Dense initial conditions
displayed in Fig.\ \ref{fig:eight_2} and a symmetric copy of those,
shifted  to the other well; unoccupied initial conditions indicated
by small open circles; the initially occupied Gaussian is displayed
by a filled circle. (b) Comparison of numerical results for
the absolute value of the cross-correlation function, using a grid of $M=98$ initial conditions shown
in panel (a) in the CCS case (solid line) and converged split
operator FFT calculations (dotted line). Both lines are
almost indistinguishable.}
\label{fig:eight_3}
\end{figure}

Secondly, to see if CCS can do better, we now launch a copy
of the grid on the right side also on the left side, as
displayed in panel (a) of Fig.\ \ref{fig:eight_3}. Both
sets of trajectories are uncoupled, as classical mechanics is a
local theory, but the coefficients
corresponding to the trajectories are coupled. Thus, although
all $a$-coefficients on the left are initially zero, they turn
non-zero in
the course of time and probability density is transferred
to the other side, thus describing tunneling almost perfectly
as can be seen in panel (b) of Fig.\ \ref{fig:eight_3},
where we have extended the time interval, such that two complete periods of tunneling can be observed.

\begin{figure}
\includegraphics[width=0.46\columnwidth,trim = 0cm 4.6cm 0cm 5cm, clip]{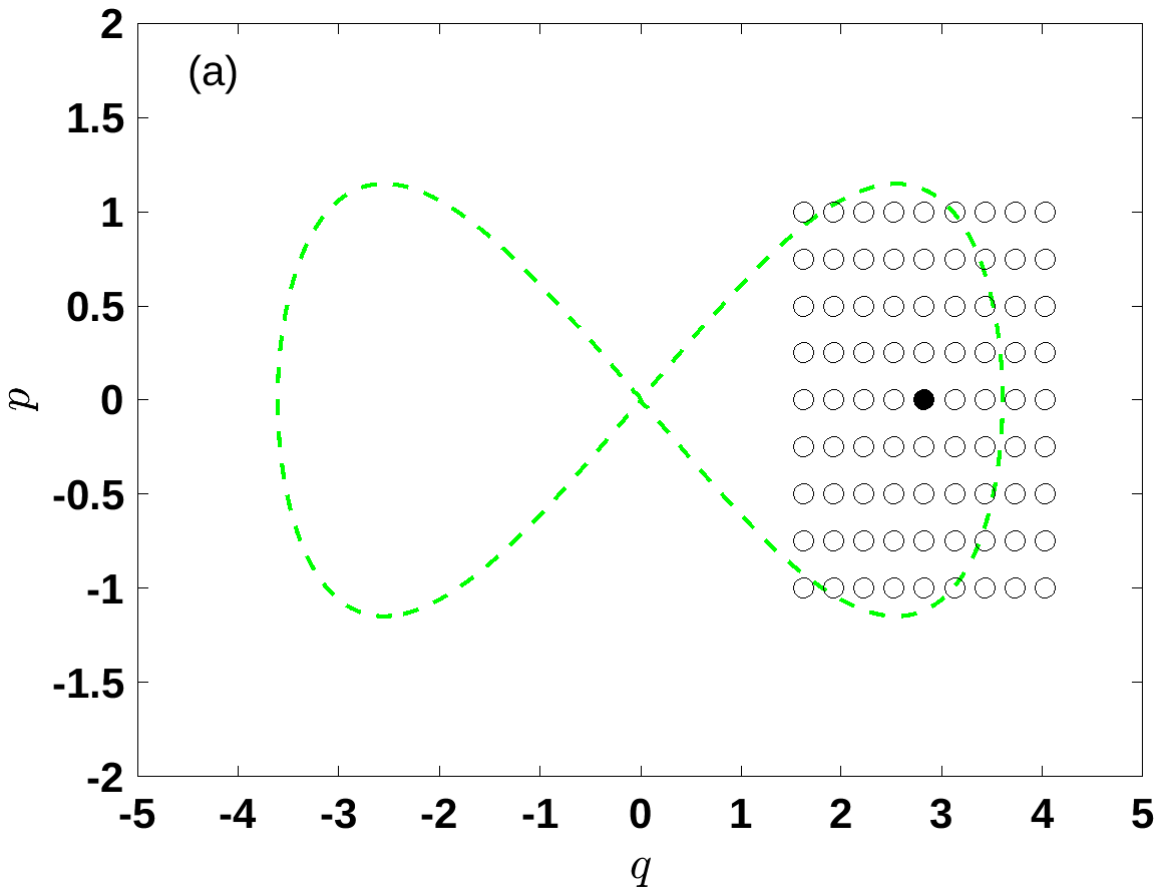}
\includegraphics[width=0.49\columnwidth,trim = 0cm 5cm 0cm 5cm, clip]{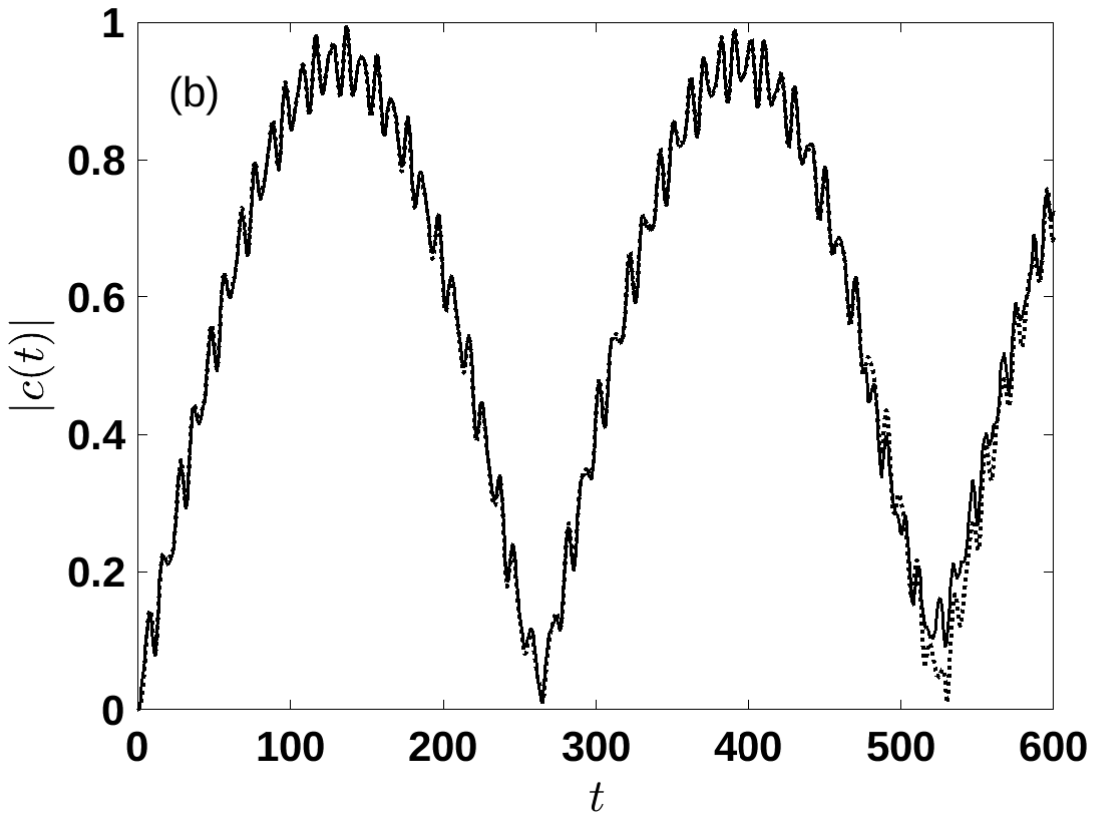}
\caption{(a) Separatrix of $H_{\rm ord}$ in the phase space of the
quartic double well with $D=1$. Less dense initial conditions
as compared to Fig.\ \ref{fig:eight_2}, unoccupied initial
conditions indicated by small open circles; the initially occupied
Gaussian is displayed by a filled circle.
(b) Comparison of numerical results for the absolute value of the
cross-correlation function, using a grid of $M=81$ initial
conditions shown in panel (a) in the CCS case (solid line) and
converged split operator FFT calculations (dotted line).}
\label{fig:eight_4}
\end{figure}

As a third try, let us increase the extension of the grid,
initially on the right side of the barrier, see panel (a) of
Fig.\ \ref{fig:eight_4}. This grid has now 81 points and an
extension which is four times as large as the one in panel (a)
of Fig.\ \ref{fig:eight_2}, leading to a density still far larger
than the von Neumann limit \cite{vonNeu}, but also leading to
a substantial part of initial conditions with energy above the
barrier, i.e., outside of the ``eight''. The comparison of
the autocorrelations in panel (b) of Fig.\ \ref{fig:eight_4},
at least for short times, is
now almost as favorable as in Fig.\ \ref{fig:eight_3}. We have not
tried to fully converge the results, to still see a little
difference between the results plotted in the graph. By retaining
the extension of the grid but allowing for 121 instead of 81
trajectories, the quality of agreement with the split operator
FFT results is almost as good as in Fig.\ \ref{fig:eight_3} (not
shown). Thus also high energy trajectories ``mimic'' the
tunneling, which happens at low energies quantum mechanically,
where there are no classical trajectories (on the left side) by our
construction.

\begin{figure}
\includegraphics[width=0.99\columnwidth,trim = 0cm 5cm 0cm 5cm, clip]{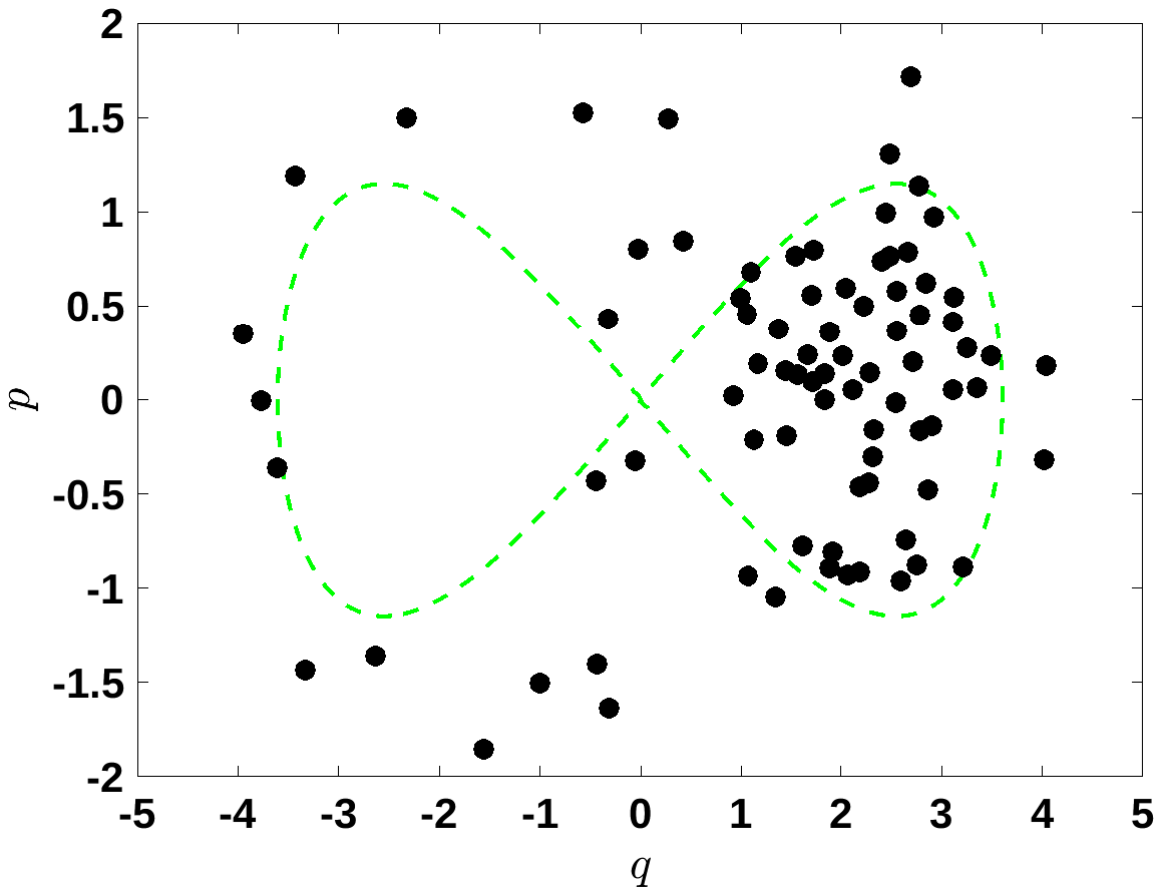}
\caption{Separatrix of $H_{\rm ord}$ in the phase space of the quartic double well with $D=1$. Snapshot of the time-evolved  less dense initial conditions panel (a) of Fig.\ \ref{fig:eight_4} at time $t=131$ (corresponding to half the tunneling period).}
\label{fig:eight_5}
\end{figure}

This fact is demonstrated in Fig.\ \ref{fig:eight_5}, where
the phase space eight, together with the trajectories at time
$t=131$, close to half the tunneling period is shown. The trajectories outside of the eight are capable of creating
a localized wavepacket at the bottom of the left well (because
the overlap with the state $|\beta\rangle$ at the considered
time is almost unity.
We do not display the relative weight of the trajectories
by the size of the dots because, due to the non-orthogonality
of the coherent states, the $a$-coefficients are not ``normalized''
in the same way as the ones in an orthogonal basis expansion
and some of them can become exceedingly large.
The norm of the wavefunction is conserved to within around
one percent for the times displayed in
panel (b) of Fig.\ \ref{fig:eight_4}, however.

\section{Summary and Outlook}

We have shown that there are two ways to accurately describe tunneling
in a double well by the use of coupled Gaussian basis functions,
moving along classical trajectories: (i) either low energy initial conditions
for classical trajectories are also launched in the
initially unpopulated well, or (ii) one
allows for trajectories with enough energy to
overcome the barrier. A combination of the two cases has already been studied long ago in \cite{SC01}.
Here we have disentangled the two cases.
Due to the fact that the equations
of motion for the coefficients are coupled, in the course
of time, the $a$-coefficients of all trajectories
will start to deviate from zero. Probability density
can thus build up in the other well by two scenarios.
Either the trajectories have been seeded in the initially unpopulated well from the very start, or they move there by having
enough energy. The use of rectangular grids is not mandatory.
We believe that also circular grids will do
the job \cite{Schl}.

Furthermore, also the investigation of incoherent barrier
tunneling is a worthwhile topic. In \cite{cpl95}, it has, e.g.,
been shown that employing classical trajectories in a purely
semiclassical (van Vleck-type) study does not lead
to converged results for the barrier transmission as a
function of the initial wavepacket center, while in \cite{prl00},
the use of multiple spawning of new classical trajectories
after time-slicing was shown to lead to
converged results. In future work, the use
of trajectories that start on one side of the barrier with
enough energy to overcome it classically in a CCS study
shall be contrasted with the use of trajectories starting on
both sides of the barrier. In addition, the investigation of
dynamical tunneling \cite{DaHe81} in the light of our present findings is left for future research.

Finally, a lot of work on higher-dimensional tunneling
applications using Gaussian wavepackets has
been performed \cite{WB04,SSC06,SH17,DWG20}, trying to reduce the number
of basis functions as well as to overcome potential
problems like zero-point energy leakage of the underlying classical trajectories. Although even in a HK approach it has been
shown that the final result is not plagued by zero-point energy leakage \cite{cp18}, for the numerics this might be an issue.
The results presented here may give further guidance in the
search for efficient numerical algorithms also for more than a
single degree of freedom.

\section*{Acknowledgements}

FG would like to thank Profs.\ D.\ Shalashilin
and S.\ Garashchuk for fruitful discussions on
Gaussian-based variational methods.

\begin{appendix}

\end{appendix}

\section*{References}

\bibliographystyle{iopart-num}

\begin{thebibliography}{10}
\expandafter\ifx\csname url\endcsname\relax
  \def\url#1{{\tt #1}}\fi
\expandafter\ifx\csname urlprefix\endcsname\relax\def\urlprefix{URL }\fi
\providecommand{\eprint}[2][]{\url{#2}}

\bibitem{BJWM00}
Beck M, Jaeckle A, Worth G and Meyer H~D 2000 {\em Phys. Rep.\/} {\bf 324} 1 --
  105

\bibitem{Lubi}
Lubich C 2008 {\em {\it From Quantum to Classical Molecular Dynamics: {R}educed
  Models and Numerical Analysis}\/} Zurich Lectures in Advanced Mathematics
  (Z\"urich: European Mathematical Society)

\bibitem{Richings2015}
Richings G~W, Polyak I, Spinlove K~E, Worth G~A, Burghardt I and Lasorne B 2015
  {\em Int. Rev. in Phys. Chem.\/} {\bf 34} 269--308

\bibitem{irpc21}
Werther M, {Loho Choudhury} S and Grossmann F 2021 {\em Int. Rev. in Phys.
  Chem.\/} {\bf 40} 81

\bibitem{KS81}
Kramer P and Saraceno M 1981 {\em Geometry of the time-dependent variational
  principle in quantum mechanics\/} (Berlin: Springer Verlag)

\bibitem{Hetal11}
Haegeman J, Cirac J~I, Osborne T~J, Pizorn I, Verschelde H and Verstraete F
  2011 {\em Phys. Rev. Lett.\/} {\bf 107}(7) 070601

\bibitem{Scholl2011}
Schollw\"ock U 2011 {\em Annals of Physics\/} {\bf 326} 96 -- 192

\bibitem{Glauber}
Glauber R~J 1963 {\em Phys. Rev.\/} {\bf 131}(6) 2766--2788

\bibitem{HK84}
Herman M~F and Kluk E 1984 {\em Chem. Phys.\/} {\bf 91} 27

\bibitem{WaHe09}
Wang Z~X and Heller E~J 2009 {\em J. Phys. A\/} {\bf 42} 285304

\bibitem{jpa16}
Ray S, Ostmann P, Simon L, Grossmann F and Strunz W~T 2016 {\em J. Phys. A\/}
  {\bf 49} 165303

\bibitem{SiSt14}
Simon L and Strunz W~T 2014 {\em Phys. Rev. A\/} {\bf 89}(5) 052112

\bibitem{Letal19}
Lando G~M, Vallejos R~O, Ingold G~L and de~Almeida A~M~O 2019 {\em Phys. Rev.
  A\/} {\bf 99}(4) 042125

\bibitem{SC01}
Shalashilin D~V and Child M~S 2001 {\em J. Chem. Phys.\/} {\bf 114} 9296

\bibitem{SC00}
Shalashilin D~V and Child M~S 2000 {\em J. Chem. Phys.\/} {\bf 113} 10028

\bibitem{Hu27}
Hund F 1927 {\em Z. Phys.\/} {\bf 43} 805--826

\bibitem{Ku72}
Kurkij\"arvi J 1972 {\em Phys. Rev. B\/} {\bf 6} 832

\bibitem{Ketal08}
Kierig E, Schnorrberger U, Schietinger A, Tomkovic J and Oberthaler M~K 2008
  {\em Phys. Rev. Lett.\/} {\bf 100} 190405

\bibitem{zpb91}
Grossmann F, Jung P, Dittrich T and H\"anggi P 1991 {\em Zeitschrift f\"ur
  Physik B\/} {\bf 84} 315

\bibitem{Reichl}
Reichl L~E 2004 {\em The Transition to Chaos: Conservative Classical Systems
  and Quantum Manifestations\/} 2nd ed (New York: Springer)

\bibitem{DPM20}
Dittrich T and {Pena Mart\'inez} S 2020 {\em Entropy\/} {\bf 22} 1046

\bibitem{pra22_1}
{Loho Choudhury} S and Grossmann F 2022 {\em Phys. Rev. A\/} {\bf 105} 022201

\bibitem{ShBu08}
Shalashilin D~V and Burghardt I 2008 {\em J. Chem. Phys.\/} {\bf 129} 084104

\bibitem{Gross3}
Grossmann F 2018 {\em Theoretical Femtosecond Physics: {A}toms and {M}olecules
  in Strong Laser Fields\/} 3rd ed (Springer International Publishing AG)

\bibitem{vonNeu}
von Neumann J 1955 {\em {\it Mathematical Foundations of Quantum Mechanics}\/}
  (Princeton: Princeton University Press)

\bibitem{Schl}
Schleich W~P 2001 {\em Quantum Optics in Phase Space\/} (Berlin: Wiley-VCH)

\bibitem{cpl95}
Grossmann F and Heller E~J 1995 {\em Chem. Phys. Lett.\/} {\bf 241} 45

\bibitem{prl00}
Grossmann F 2000 {\em Phys. Rev. Lett.\/} {\bf 85} 903

\bibitem{DaHe81}
Davis M~J and Heller E~J 1981 {\em The Journal of Chemical Physics\/} {\bf 75}
  246--254

\bibitem{WB04}
Wu Y and Batista V~S 2004 {\em J. Chem. Phys.\/} {\bf 121} 1676--80

\bibitem{SSC06}
Sherratt P~A, Shalashillin D~V and Child M~S 2006 {\em Chemical Physics\/} {\bf
  322} 127--134

\bibitem{SH17}
Saller M~A~C and Habershon S 2017 {\em J. Chem. Theory Comput.\/} {\bf 13} 3085

\bibitem{DWG20}
Dutra M, Wickramasinghe S and Garashchuk S 2020 {\em J. Phys. Chem. A\/} {\bf
  124} 9314--9325

\bibitem{cp18}
Buchholz M, Fallacara E, Gottwald F, Ceotto M, Grossmann F and Ivanov S~D 2018
  {\em Chem. Phys.\/} {\bf 515} 231

\end{thebibliography}
\providecommand{\newblock}{}

\end{document}